\documentclass[prl,aps,preprint,showpacs]{revtex4-1}
\usepackage{graphicx}
\usepackage{amsmath}
\usepackage[table]{xcolor}

\begin{document}

\title{Large-Gap Quantum Spin Hall States in Stanene Grown on Substrate}

\author{Yong \surname{Xu}$^{1,2}$}
\author{Peizhe \surname{Tang}$^{1}$}
\author{Shou-Cheng \surname{Zhang}$^{1,2}$}

\affiliation{$^1$Department of Physics, McCullough Building, Stanford University, Stanford, California 94305-4045, USA \\
$^2$Department of Physics and Institute for Advanced Study, Tsinghua University, Beijing 100084, People's Republic of China}

\begin{abstract}
Two-dimensional stanene is a promising candidate material for realizing room-temperature quantum spin Hall (QSH) effect. Monolayer stanene has recently been fabricated by molecular beam epitaxy, but shows metallic features on Bi$_2$Te$_3$(111) substrate, which motivates us to study the important influence of substrate. Based on first-principles calculations, we find that varying substrate conditions considerably tunes electronic properties of stanene. The supported stanene gives either trivial or QSH states, with significant Rashba splitting induced by inversion asymmetry. More importantly, large-gap (up to 0.3 eV) QSH states are realizable when growing stanene on various substrates, like the anion-terminated (111) surfaces of SrTe, PbTe, BaSe and BaTe. These findings provide significant guidance for future research of stanene and large-gap QSH states.  
\end{abstract}


\maketitle

One of the grand challenge problems in condensed matter physics and material science is to develop room-temperature electron
conduction without dissipation. Quantum spin Hall (QSH) insulators are new states of quantum matter with an insulating bulk and metallic edge states~\cite{hasan2010,qi2011}. Their helical edge states are protected against backscattering by time-reversal symmetry, enabling dissipationless conduction~\cite{hasan2010,qi2011}. However, the working temperature of QSH insulators in experiment (like HgTe quantum well~\cite{bernevig2006,konig2007}) is very low (below 10 K), limited by their small energy gap. The search of QSH insulators of large gaps becomes critically important.

A few large-gap QSH insulators have been theoretically proposed~\cite{xu2013,qian2014,weng2014,tang2014,zhou2014,si2014,song2014,zhou2014nl,chuang2014,jiang2014,luo2015,ma2015}, predominantly in ultrathin materials of free-standing form. Among them, stanene, the graphene counterpart of tin (Sn), is of special interest~\cite{xu2013}. Stanene belongs to the important class of two-dimensional (2D) group-IV materials that have attracted extensive research interest in the recent years. In contrast to other lighter materials (including graphene, silicene and germanene), stanene is characterized by strong spin-orbit coupling (SOC). The decorated stanene supports large-gap QSH state~\cite{xu2013} as well as many other novel features, like enhanced thermoelectric performance~\cite{xu2014}, near room temperature quantum anomalous Hall (QAH) effect~\cite{wu2014} and novel topological superconductivity~\cite{wang2014}. These novel properties make stanene systems promising for fundamental research and future technologies.

The research of stanene is a fast-growing field. Recently, monolayer stanene has been successfully fabricated by molecular beam epitaxy (MBE) on Bi$_2$Te$_3$(111) substrate~\cite{zhu2015}, which, for the first time, established the existence of the theoretically predicted stanene structure. Unfortunately, valence bands of stanene are pinned with conduction bands of Bi$_2$Te$_3$(111), giving metallic interface states~\cite{zhu2015}. When growing stanene on a different substrate, the binding configuration and bonding strength vary, and the charge redistribution and energy level alignment at the interface also differ. The combined effects would lead to distinct electronic structures for ultrathin materials like stanene.  The important influence of substrate, however, has been largely ignored in previous works~\cite{xu2013,qian2014,weng2014,tang2014,zhou2014,si2014,song2014,zhou2014nl,chuang2014,jiang2014,luo2015,ma2015}. Realizing large-gap QSH states in ultrathin materials grown on substrate emerges as a challenging issue.

In fact, substrate could provide additional degrees of freedom, like strain and magnetism, to tune properties of the supported 2D material. Desirable material features thus can be designed by engineering the substrate. To demonstrate the concept, we study properties of the supported stanene on various substrates by first-principles calculations. The results reveal that the substrate plays a crucial role in determining electronic structure of stanene. The supported stanene can be either trivial or QSH insulator, dependent on the termination, lattice strain and chemical composition of the substrate. Other key electronic characteristics, like band gap and Rashba splitting, can also be controlled by varying substrate conditions. Crucially, QSH states of large gaps (up to 0.3 eV) can be obtained when growing stanene on a series of substrates, like the anion-terminated (111) surfaces of SrTe, PbTe, BaSe and BaTe. The findings shed light on future research of stanene and large-gap QSH states.

First-principle calculations were performed by using density functional theory (DFT) methods as implemented in the Vienna \textit{ab initio} simulation package~\cite{kresse1996}. The projector-augmented-wave potential, the Perdew-Burke-Ernzerhof exchange-correlation functional~\cite{perdew1996} and the plane wave basis with an energy cutoff of 250 eV were employed. Stanene on substrate was modeled by periodic slab approach including a vacuum layer of over 13 \AA \ together with dipole corrections between periodic images, using 11$\times$11$\times$1 Monkhorst-Pack $k$ grid. Note that asymmetric slab model includes an unwanted bottom surface that even being saturated may introduce artificial surface states within the bulk gap. To avoid the disadvantage, symmetric slab model was applied, including a 13-atomic-layer substrate material with the middle 7 layers fixed at the bulk crystal structure.  The two symmetric surfaces were covered by two copies of stanene. Only the top half part of the slab model was displayed in structure figures. The experimental lattice constants were used for bulk PbTe~\cite{guergouri1993}, SrTe~\cite{kobler1993}, BaSe~\cite{miller1960} and BaTe~\cite{miller1960}, corresponding to surface lattice constants of 4.568, 4.711, 4.667 and 4.953 \AA, respectively, for their (111) surfaces. The force convergence criterion of 0.01 eV$/$\AA \ was chosen for structural optimization. The SOC was included in self-consistent electronic structure calculations. The maximally localized Wannier functions~\cite{marzari1997} constructed from DFT were used to calculate edge states.

Stanene is an atomically thin sheet of Sn atoms in a buckled honeycomb lattice (see Fig. 1) . In free-standing stanene, the $\pi$-$\pi$ bonding between $p_z$ orbitals of Sn atoms forms a Dirac cone at the $K$ ($K'$) point when excluding the SOC, and the inclusion of the SOC opens a nontrivial band gap. The system is a QSH insulator similar to graphene~\cite{kane2005} but having a much larger band gap ($\sim$0.1 eV) because of the stronger SOC~\cite{xu2013}. One important characteristic of the free-standing stanene is that its unsaturated $p_z$ orbital is chemically active, due to the weak $\pi$-$\pi$ interaction as caused by the large bond length between Sn atoms ($\sim$3 \AA).  This feature together with the out-of-plane orientation of the $p_z$ orbital facilitate strong orbital interaction with external environments. Therefore, electronic properties of bare stanene are easily affected by adsorbates and substrate~\cite{zhu2015}. The disadvantage may be overcome by passivating the $p_z$ orbital.

A desirable substrate for growing stanene with observable QSH states is expected to satisfy following conditions: (i) It has an insulating bulk gap. (ii) The surface Bravais lattice is hexagonal, commensurate with that of stanene, and the lattice mismatch is small. (iii) The coupling with stanene is strong enough to enable layer-by-layer growth of monolayer stanene. Many semiconducting or insulating substrates meet the requirements, like (111) surfaces of rhombohedral and cubic lattices and (001) surfaces of hexagonal lattices. The large amount of candidate substrates offers great possibilities for tuning properties of the supported stanene. We will focus on (111) surfaces of rock-salt crystals $AB$ ($A$ = Pb, Sr, Ba and $B$ = Se, Te), for which high quality substrates with atomically flat terraces are experimentally available~\cite{yan2014}. On these substrates, the $p_z$ orbital of stanene couples strongly with the dangling bond of the substrate, largely stabilizing the whole system. Layer-by-layer growth of monolayer stanene thus is possible, for instance, by MBE.  To make stanene inert to adsorbates, $p_z$ orbitals of the top Sn atoms exposed to vacuum are passivated by chemical functional groups $X$ as shown in Fig. 1.

\begin{figure*}
\includegraphics[width=\linewidth]{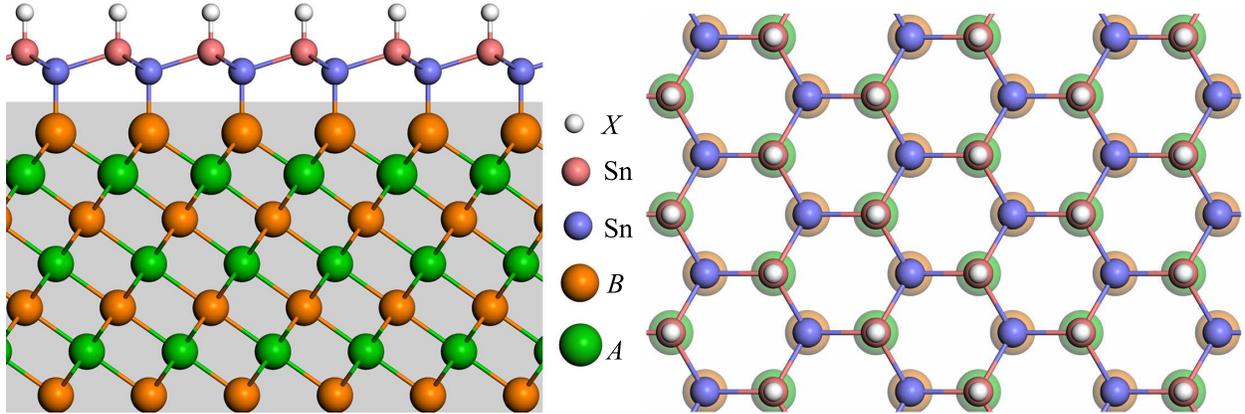}
\caption{Crystal structure for stanene grown on $AB$(111)-$B$ from the side (top) view [left (right)], where $A$ = Pb, Sr, Ba and $B$ = Se, Te. Stanene is an atomically thin sheet of Sn atoms in a buckled honeycomb lattice, for which the top and bottom Sn atoms are represented by balls of different colors. The top Sn atoms of stanene are decorated by chemical functional groups $X$ ($X = $ -H, -F, -Cl, -Br, -I, etc.) to saturate the $p_z$ orbital.}
\end{figure*}

Let's first look at an example substrate of SrTe(111)-Te. To search stable structures, our first-principles calculations started from many initial geometries, obtained several locally stable geometries by structural relaxation, and then selected the ones with lowest total energies. We observe the same most stable binding site for the supported stanene without and with its top Sn atoms decorated by iodine (i.e., $X$ = -I), and will only discuss the later case. In the most stable configuration (see Fig. 1), the bottom (top) Sn atoms of stanene are located at the top (hcp) sites of the substrate, denoted as ``top-hcp'' for simplicity. The ``top-fcc'' configuration shares close total energy (0.01 eV larger) and similar electronic structure. On this Te-terminated surface, Sn atoms binds preferably on top of Te atoms by forming chemical bonds, so that the $p_z$ orbital of stanene and the dangling bond of the substrate both get fully passivated. In contrast, other metastable configurations, like ``hcp-top'' and ``hcp-fcc'', are much less stable, over 0.8 eV higher in energies. Their band structure could be metallic or semiconducting, which is rather system dependent.

\begin{figure*}
\includegraphics[width=\linewidth]{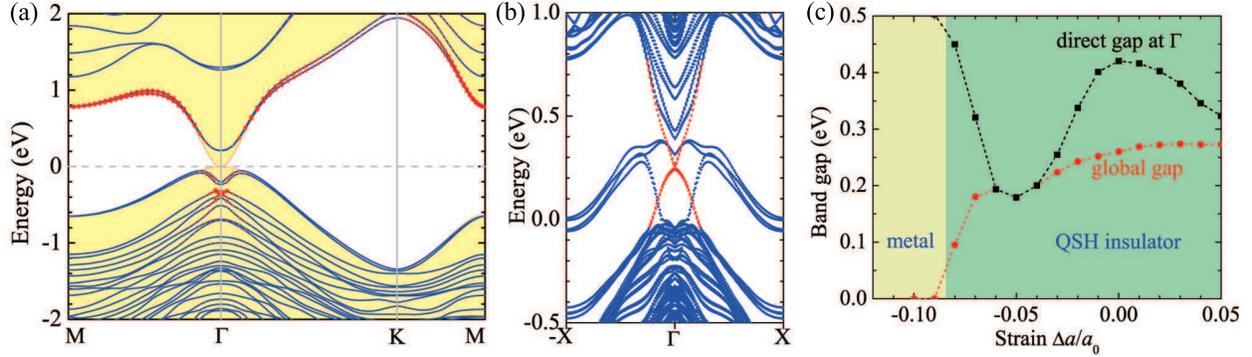}
\caption{Electronic structure for stanene grown on SrTe(111)-Te with the top Sn atoms iodinated ($X = $ -I, see geometry in Fig. 1). (a) Band structure including (excluding) the spin-orbit coupling, denoted by blue lines (shaded regions). The red dots on bands represent the contribution from the $s$ orbital of stanene, which indicates a band inversion at the $\Gamma$ point. (b) Band structure of a zigzag ribbon structure showing the existence of gapless helical edges (red colored). (c) Band gaps as a function of lattice strain $\Delta a/ a_0$, where $a_0$ is the equilibrium lattice constant of substrate and $\Delta a$ is the change of lattice constant. The supported stanene keeps as QSH insulator when under moderate strain conditions.}
\end{figure*}

Figure 2(a) presents the calculated band structure for the most sable configuration. Around the Fermi level, a few bands appear within the bulk gap of substrate. They are mostly contributed by Sn atoms according to wavefunction analysis, thus identified as ``stanene-related'' bands. A type-I energy level alignment is formed between stanene and substrate. For ``stanene-related'' bands, a large band gap is opened at the $K$ ($K'$) point due to the saturation of the $p_z$ orbital of stanene.  The low-energy physics occurs around the $\Gamma$ point, where the band structure is gapless without the SOC and becomes gapped with the SOC.

Near the $\Gamma$ point, the relevant ``stanene-related'' energy levels are given by $|s \rangle$, $|p_{x,y} \rangle$, $|s^* \rangle$ and $|p_{x,y}^* \rangle$, which correspond to bonding and anti-bonding states of  $s$ and $p_{x,y}$ orbitals of Sn atoms, respectively. These energy levels are filled by six $s$ and $p_{x,y}$ electrons, considering that there are two Sn atoms in one unit cell and their $p_{z}$ orbitals are passivated. In trivial insulators, like the carbon and silicon counterparts, $|s^* \rangle$ is higher in energy than $|p_{x,y} \rangle$ at the $\Gamma$ point. Then $|s \rangle$ and $|p_{x,y} \rangle$ are occupied, leaving $|p_{x,y} \rangle$ as the valence band maximum (VBM) and $|s^* \rangle$ as the conduction band minimum (CBM). A band inversion between bonding $|p_{x,y} \rangle$ and anti-bonding $|s^* \rangle$  can result in nontrivial QSH phase, as learned from HgTe quantum well~\cite{bernevig2006,konig2007} and free-standing decorated stanene~\cite{xu2013}. The same mechanism also works here for the supported stanene as to be shown below.

To check the band inversion, we project Bloch wavefunctions onto atomic orbitals of stanene. Bloch states contributed by $s$ orbital of Sn (or $|s^* \rangle$ of stanene) are visualized in Fig. 2(a) by red dots. These states stay above the Fermi level for most of Brillouine zone, but shift to valence bands around the $\Gamma$ point. The CBM is given by $|p_{x,y} \rangle$ of stanene, instead of $|s^* \rangle$ for usual cases. A band inversion between $|s^* \rangle$ and $|p_{x,y} \rangle$ is identified at the $\Gamma$ point. Then the occupation of energy levels at the $\Gamma$ point is determined:  the lower $|s \rangle$ and $|s^* \rangle$ are fully occupied, leaving two electrons on the higher $|p_{x,y} \rangle$. Without the SOC the Fermi level would cross the degenerate $|p_{x,y} \rangle$ bands, leading to zero band gap, and the inclusion of SOC opens a band gap, as observed by our calculations. This SOC-induced band gap is topologically nontrivial as explained by the band inversion, thus the supported stanene is a QSH insulator.

As an hallmark of QSH insulator, gapless helical edge states with spin-momentum locking are topologically protected to appear within the bulk gap. By using the maximally localized Wannier functions~\cite{marzari1997}, we construct the \textit{ab initio} tight binding model for the zigzag nanoribbon with the width of 15 unit cells. The calculated electronic states are shown in Fig. 2(b): despite some trivial edge states induced by dangling bonds, nontrivial gapless bands (highlighted by red color) exist within the bulk gap, connecting bulk valence bands with bulk conduction bands. These edge states are spin-polarized, with opposite group velocities for opposite edges. The existence of gapless helical edge states together with the analysis of band inversion consistently prove that the supported stanene is a QSH insulator.

Noticeably,  a nontrivial band gap of $0.26$ eV is obtained, which is sufficiently large to support room-temperature QSH effect. One advantage of the present system is that its QSH states can be tuned by varying various conditions, like the chemical functional group $X$ or the lattice strain of the substrate. We replace $X=$ -I by  -F, -Cl, -Br and -H, and find that the band gaps are all nontrivial and keep large, which are 0.22, 0.21, 0.20 and 0.17 eV, respectively. Other functional groups, like -OH, -CH$_3$, -CN, -C$_2$H, -NH$_2$ etc are not considered here, but can also be used to modify QSH states of stanene. Moreover, as shown in Fig. 2(c), when varying the lattice constant of the substrate, the support stanene keeps as QSH insulator at most strain conditions and becomes metallic only at very large compressive strain of over $8\%$. When the lattice is compressed, the nontrivial band gap transits from indirect to direct and back to indirect, with its value tunable in a wide range, from zero to $\sim$0.3 eV. Experimentally, the lattice constant of substrate can be controlled, for instance, by alloying SrTe with other materials like PbTe and BaTe.

Rashba splitting is well known to be important for spintronics, and is essential to realize novel 2D topological superconductivity in stanene~\cite{wang2014}. There is no Rashba splitting in the free-standing stanene because of the inversion symmetry. In contrast, when coupling stanene with the substrate, the inversion symmetry is naturally broken, introducing Rashba splitting [see Fig. 2(a)]. Since the strength of inversion asymmetry can be controlled by varying either the substrate conditions or the chemical functional group, Rashba splitting in principle becomes tunable in the supported stanene, which is advantageous for spintronics and topological superconductivity.

\begin{figure*}
\includegraphics[width=\linewidth]{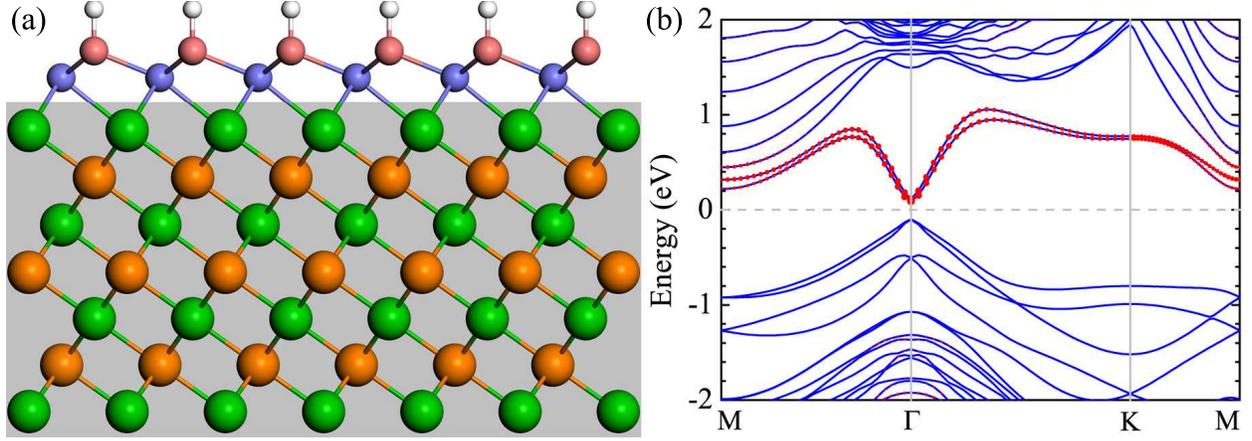}
\caption{(a) Crystal structure for stanene grown on $AB$(111)-$A$ from the side view (using the same notations as in Fig. 1). (b) Band structure for stanene grown on SrTe(111)-Sr with the top Sn atoms iodinated. The red dots on bands represent the contribution from the $s$ orbital of stanene, which indicates no band inversion at the $\Gamma$ point. }
\end{figure*}

We have demonstrated above that large-gap QSH states can be realized by growing stanene on SrTe(111)-Te. However, the geometry, electronic structure and topological nature of the supported stanene could change dramatically when varying the substrate termination. On SrTe(111)-Sr, the bottom Sn atoms of stanene preferably bind at the fcc site, different from the top site on the Te-terminated substrate. The most stable binding configuration of stanene becomes ``fcc-hcp''  [see Fig. 3(a)].  The different binding geometry is presumably explained by the distinct bonding nature of Sn-Sr and Sn-Te bonds, as the later is more like valence bond than the former. In addition to the variance of geometry, electronic structure changes noticeably. As shown in Fig. 3(b), bands contributed by $|s^* \rangle$ of stanene stay above the Fermi level in the whole Brillouine zone. No band inversion between $|s^* \rangle$  and $|p_{x,y} \rangle$ occurs at the $\Gamma$ point, suggesting that the supported stanene becomes topologically trivial. Interestingly, the binding of stanene on SrTe(111)-Sr induces significant electron transfer from Sr to Sn and from Sn to the halide group, which builds up a strong electric field across stanene, inducing large Rashba splitting.

Having understood properties of stanene grown on SrTe(111), we vary the chemical composition of substrate and consider other $AB$(111) substrates, including PbTe, BaSe and BaTe. The results are qualitatively similar to those of SrTe. On $AB$(111)-$A$, the most stable binding configuration of stanene is ``fcc-hcp'' [see Fig. 3(a)], and the supported stanene is topologically trivial, which is metallic on PbTe and semiconducting on BaSe and BaTe (data not shown). In contrast, on $AB$(111)-$B$, the most stable binding configuration is ``top-hcp'' (see Fig. 1). Their band structures are all characterized by a band inversion at the $\Gamma$ point (see Fig. 4). The supported stanene is QSH insulator for all these substrates of PbTe, BaSe and BaTe, whose nontrival band gaps are 0.09, 0.24, and 0.30 eV, respectively. Sizable QSH gap of 0.3 eV is obtained, which is useful for low-dissipation electronics. Moreover, QSH states can be achieved in the supported stanene for this series of substrates. Therefore, its properties can be systematically studied and be continuously tuned by alloying the substrate materials.

\begin{figure*}
\includegraphics[width=\linewidth]{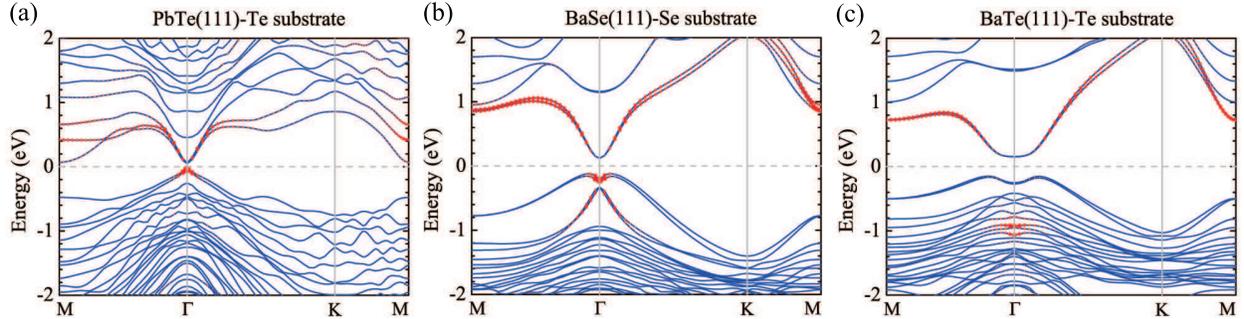}
\caption{Band structures for stanene grown on (a) PbTe(111)-Te, (b) BaSe(111)-Se and (c) BaTe(111)-Te, which give nontrivial QSH gaps of 0.09, 0.24, and 0.30 eV, respectively (see geometry in Fig. 1 with $X = $ -I).  The red dots on bands represent the contribution from the $s$ orbital of stanene, which indicates a band inversion at the $\Gamma$ point.}
\end{figure*}

In summary, we show that properties of the supported stanene can be greatly tuned by varying the substrate conditions, based on the discussion of various candidate substrates, including (111) surfaces of SrTe, PbTe, BaSe and BaTe. Topological trivial states are obtained in stanene grown on the cation-terminated substrate. In contrast, QSH states of large gaps (up to 0.3 eV) are obtained when using the anion-terminated substrate, for which type-I energy level alignment~\cite{einspruch2014heterostructures} is formed between stanene and substrate, and the nontrivial gap is induced by band inversion between bonding and anti-bonding states of stanene itself. Generally, when growing stanene on other substrates, it is possible to get type-II and type-III energy level alignments~\cite{einspruch2014heterostructures} as well. Band inversion may occur between stanene states and substrate states~\cite{tang2014}, which could also induce QSH states. Furthermore, the use of magnetic substrates, like EuTe(111), may induce band inversion for one spin channel only, giving rise to novel QAH states. Moreover, the appearance of Rashba splitting induced by inversion asymmetry of the supported stanene could help realize novel 2D topological superconductivity. In this context, the subject of stanene grown on substrate represents a new research direction for studying novel topological states and low-dimensional physics.

This work was supported in part by FAME, one of six centers of STARnet, a Semiconductor Research Corporation program sponsored by MARCO and DARPA and by the NSF under grant numbers DMR-1305677.

%

\end{document}